\documentclass[a4paper,11pt]{article}
\pdfoutput=1 

\usepackage{jheppub} 

\usepackage[T1]{fontenc} 
\usepackage{tikz-cd}
\usepackage{caption}

\title{\boldmath  Exact Stress Tensor flows in Causal Electrodynamics: Courant-Hilbert Approach}

\author[b]{Zahra Pezhman,}
\author[b]{Komeil Babaei Velni,}
\author[a]{H. Babaei-Aghbolagh}

\affiliation[a]{Institute of Fundamental Physics and Quantum Technology, \& School of Physical Science and Technology, Ningbo University, Ningbo, Zhejiang 315211, China}
\affiliation[b]{Department of Physics, University of Guilan, P.O. Box 41335-1914, Rasht, Iran}
\emailAdd{ zpezhmanh@phd.guilan.ac.ir}
\emailAdd{hosseinbabaei@nbu.edu.cn}
\emailAdd{babaeivelni@guilan.ac.ir}

\abstract{It has been shown that for theories whose Courant-Hilbert function depends on a single dimensionful coupling, the energy-momentum tensor is controlled by the two combinations $\tau\dot{\ell}$ and $\ell-\tau\dot{\ell}$, so that the quadratic invariants $X=T_{\mu\nu}T^{\mu\nu}$ and $Y=(T^\mu{}_\mu)^2$ depend on a single auxiliary variable. Eliminating this variable in favour of $Y/X$ yields
the exact autonomous flow $\partial_\lambda\mathcal L=X\,\mathcal F(Y/X)$,
with no explicit dependence on the coupling, the electromagnetic
invariants or the CH parameter, on every branch where the map is
invertible. The expansion of $\mathcal F$ in powers of $\sqrt{Y/X}$
shows which flows are built from integer powers of $X$ and $Y$ and which
require half-integer powers. In this paper, we work out four examples in closed form, each with its analytic structure and physical interpretation: the generalized Born-Infeld theory, whose flow is quadratic in the stress tensor; the $q$-deformed family, in which the odd coefficients of the expansion vanish at $q=1/2$ and are nonzero otherwise; the no-$\tau$-maximum theory; and the causal Russo-Townsend logarithmic theory, whose flow involves the lower real branch of the Lambert $W$ function and reproduces the known perturbative result. The singular points of $\mathcal{F}$ are related to the critical fields, energy conditions, and strong-field behavior of each theory.}

\begin{document}
\maketitle
\flushbottom
\section{Introduction}
\label{sec:introduction}

The space of consistent irrelevant deformations of a given quantum field
theory is, in general, poorly understood. The notable exception is the
$T\bar T$ deformation of two-dimensional integrable field
theories~\cite{Smirnov:2016lqw,Cavaglia:2016oda}, whose solvability has
led to a remarkably complete picture of the deformed spectrum, the
factorized scattering, and the underlying integrable
structure~\cite{Zamolodchikov:2004ce,Cardy:2018sdv,Bonelli:2018kik,
Conti:2018jho}. Attempts to extend this success to higher dimensions
\cite{Taylor:2018xcy,Hartman:2018tkw,Conti:2022egv} have revealed a rich
interplay between irrelevant deformations, electromagnetic duality, and
nonlinear electrodynamics~\cite{Babaei-Aghbolagh:2020kjg,Ferko:2022iru,
Ferko:2023ruw,Ferko:2019oyv}. In four dimensions, the leading example is Maxwell theory deformed by the quadratic stress-tensor operator~\cite{Babaei-Aghbolagh:2020kjg,Ferko:2022iru,Conti:2022egv}, which generates Born-Infeld electrodynamics~\cite{Born:1934gh}. When a single scalar field is present in the DBI action~\cite{Tseytlin:1999dj}, this deformation can be interpreted as a $T\bar{T}$-like deformation~\cite{Brizio:2026ynw}.
The complementary
marginal deformation, built instead from the square root of the
stress-tensor combination, generates the duality-invariant conformal
extension of Maxwell theory known as
ModMax~\cite{Bandos:2020jsw,Kosyakov:2020wxv}; in the language of
stress-tensor flows it is generated by the root-$T\bar T$
operator~\cite{Ferko:2022iru,Ferko:2022cix,Babaei-Aghbolagh:2022leo,
Borsato:2022tmu}. Reference~\cite{Ferko:2023wyi} proved that all self-dual theories (except the Bialynicki-Birula theory) arise from a stress-tensor flow off a Maxwell seed, but did not give a general closed form of the flow in terms of stress-tensor invariants. Here we present the general flow equations for a single-parameter subfamily of NEDs with two couplings, $\gamma$ and $\lambda$, and derive, for various examples, exact coupling-independent flows that depend only on the two energy-momentum-tensor structures.



A convenient framework for this question is the Courant--Hilbert (CH)
representation~\cite{10.1115/1.3630089}. For a Lagrangian that depends only on
the field strength $F_{\mu\nu}$ and its dual, electromagnetic self-duality
is equivalent to the partial differential equation
$\mathcal L_U\mathcal L_V=-1$, where $U$ and $V$ are the non-negative
combinations of the two Lorentz invariants. Its general solution is
implicit and is parametrized by a single function $\ell(\tau)$ of an
auxiliary variable
$\tau$~\cite{Gaillard:1997rt,Gaillard:1997zr,Gibbons:1995cv,
Sorokin:2021tge}; for the origin of self-duality in nonlinear
electrodynamics see~\cite{Gaillard:1981rj}. Every self-dual theory corresponds to a choice of
$\ell(\tau)$, and its energy--momentum tensor is encoded in this single
function. Causality and convexity reduce to the two simple conditions
$\dot\ell\ge1$ and $\ddot\ell\ge0$~\cite{Russo:2024llm,Russo:2024xnh}
(for causality in Born--Infeld theory see also~\cite{Gibbons:2000xe}). The CH
formulation has recently proved to be a powerful tool for studying
deformations of integrable field theories and nonlinear
electrodynamics~\cite{Ferko:2023wyi,Ferko:2024zth,
Babaei-Aghbolagh:2025uoz,Babaei-Aghbolagh:2025hlm,
Fukushima:2025tlj,Fukushima:2026gan}, and the complete CH deformation
can itself be realized as the on-shell action of a two-dimensional
massive gravity theory~\cite{Fukushima:2026znr,Baglioni:2026mii}.

In this paper we study the irrelevant flows of one-scale CH theories,
for which $\ell(\tau;\lambda,\gamma)=\lambda^{-1}f(e^\gamma\lambda\tau)$
with one dimensionful coupling $\lambda$ and the dimensionless ModMax
parameter $\gamma$. The key observation is that the energy--momentum
tensor of such a solution decomposes into a traceless Maxwell part and a
trace part, with coefficients fixed by the two combinations $\tau\dot\ell$
and $\ell-\tau\dot\ell$. As a result, the two independent quadratic
stress-tensor invariants $X=T_{\mu\nu}T^{\mu\nu}$ and
$Y=T^\mu{}_\mu T^\nu{}_\nu$ are determined by a single auxiliary
variable, and the deformation parameter can be eliminated in favour of
their ratio. This gives the autonomous form~\cite{Pezhman:2026dsr}
\begin{equation}
\frac{\partial\mathcal L}{\partial\lambda}
=
X\,\mathcal F\!\left(\frac{Y}{X}\right),
\label{intro_main}
\end{equation}
where the flow function $\mathcal F$ is fixed by $(X,Y)$ and contains no
explicit dependence on the coupling, the electromagnetic invariants, or
the auxiliary CH parameter.
Reference~\cite{Babaei-Aghbolagh:2026gbr} reconstructs exact autonomous, independent stress-tensor generators for prescribed one-scale Courant-Hilbert families in two-dimensional integrable theories.
We stress what this statement does and does not say. The flow function is specific to the family under consideration, so \eqref{intro_main} is not a single universal operator acting on all theories. It is defined on the range of the map from the CH variable to $Y/X$, and it is single-valued only on branches where this map is invertible. Within these limits, it expresses the deformation of
each family as an exact function of the stress tensor alone, without
reference to the particular parametrization of the Lagrangian.

The operator content of the flow is read off from the expansion of
$\mathcal F(Y/X)$ around the $Y \to 0$,
\begin{equation}
\frac{\partial\mathcal L}{\partial\lambda}
=
\sum_{m=0}^{\infty}
C_m\,
Y^{m/2}X^{1-m/2}.
\label{intro_expansion}
\end{equation}
In this expansion, the coefficients $C_m$ determine the exact functional form of the $T\bar{T}$ flow equation in terms of the two structures $X$ and $Y$. We obtain the closed form of this series for electrodynamics theories generated by the Courant-Hilbert approach..

We illustrate the construction with four theories, each treated in a
self-contained section. The generalized Born--Infeld theory, which
includes the ModMax coupling, has a flow that is exactly quadratic in the
stress tensor~\cite{Babaei-Aghbolagh:2020kjg,Babaei-Aghbolagh:2022uij,
Ferko:2022iru,Conti:2022egv}. The $q$-deformed family contains it as the
member $q=1/2$, and for generic $q$ the odd coefficients are non-zero. The
no-$\tau$-maximum theory, which is the member $q=3/2$ at negative
coupling, provides a closed-form flow with half-integer powers. The
causal Russo--Townsend logarithmic theory~\cite{Russo:2024ptw} (a causal
self-dual version of logarithmic electrodynamics~\cite{Soleng:1995kn}) yields an
exact flow involving the physical branch of the Lambert $W$ function~\cite{Corless:1996},
whose weak-field expansion reproduces the perturbative $T\bar T$-like
flow obtained previously~\cite{Babaei-Aghbolagh:2025cni}; the finite
self-energy of a charged particle in this theory was found
in~\cite{Babaei-Aghbolagh:2025tim}. The weak-field coefficients of the
$q$-deformed and no-$\tau$-maximum flows were obtained perturbatively
in~\cite{Babaei-Aghbolagh:2026vkm}, and we reproduce them as a check. The
new elements here are the exact autonomous form for general one-scale CH
functions, the closed-form flows of the no-$\tau$-maximum and logarithmic
theories, and the analysis of the singular points of each flow function
and of their physical meaning: critical fields, energy conditions, and
strong-field asymptotics. We also prove that the
autonomous irrelevant flow commutes exactly with the marginal root-$T\bar T$
flow $\partial_\gamma\mathcal L=\tfrac14\sqrt{4X-Y}$
(section~\ref{sec:commutativity}), which is the non-perturbative version of
the commutative diagram known from the perturbative
analysis~\cite{Babaei-Aghbolagh:2024uqp}.

The paper is organized as follows. In
section~\ref{sec:exact_flows} we review the Courant--Hilbert
formulation of self-dual nonlinear electrodynamics, derive the exact
parametric and autonomous stress-tensor flows, and discuss the operator
expansion. The same section collects the causality conditions and the
dictionary between the stress tensor and the physical variables (energy
density, pressure and the ratio $\kappa$) that is used throughout the
paper. Each of the following four sections is devoted to a single theory
and is self-contained: it contains the Courant--Hilbert function, the
exact flow, the weak-field expansion, the analytic structure and singular
points of the flow, and their physical interpretation. The theories are
the generalized Born--Infeld theory (section~\ref{sec:GBI_flow}), the
$q$-deformed family (section~\ref{sec:q_deformed}), the
no-$\tau$-maximum theory (section~\ref{sec:NM}), and the causal
Russo--Townsend logarithmic theory (section~\ref{sec:logarithmic}).
Section~\ref{sec:commutativity} shows that the irrelevant
flow derived here commutes with the root-$T\bar T$ flow. We
conclude in section~\ref{sec:conclusions} with a summary, the limitations
of the construction, and an outlook.

\section{ Exact $T\bar{T}$-flows for self-dual nonlinear electrodynamics}
\label{sec:exact_flows}
The two Lorentz invariants constructed from the electromagnetic field strength
are
\begin{equation}
S=-\frac{1}{4}F_{\mu\nu}F^{\mu\nu}
=\frac{1}{2}\left(\vec E^{\,2}-\vec B^{\,2}\right),
\qquad
P=-\frac{1}{4}F_{\mu\nu}\widetilde F^{\mu\nu}
=\vec E\cdot\vec B .
\end{equation}
It is convenient to trade $(S,P)$ for the non-negative variables
\begin{equation}
U=\frac{\sqrt{S^2+P^2}-S}{2},
\qquad
V=\frac{\sqrt{S^2+P^2}+S}{2}.
\end{equation}
In terms of these variables, the condition of electromagnetic self-duality
reduces to the Courant--Hilbert equation
\begin{equation}
\mathcal L_U\mathcal L_V=-1.
\label{CH}
\end{equation}
Its general solution can be parametrized by a single function $\ell(\tau)$,
\begin{equation}
\mathcal L
=
\ell(\tau)-\frac{2U}{\dot\ell(\tau)},
\qquad
\tau
=
V+\frac{U}{\dot\ell(\tau)^2},
\label{Lg}
\end{equation}
where $\dot\ell\equiv d\ell/d\tau\neq0$. The physical domain is
$\tau\geq0$, with $\tau=0$ at $U=V=0$. Differentiating \eqref{Lg} gives
\begin{equation}
\mathcal L_V=\dot\ell,
\qquad
\mathcal L_U=-\frac{1}{\dot\ell},
\end{equation}
so that the Courant--Hilbert equation \eqref{CH} is identically satisfied.

A useful consequence of the parametrization \eqref{Lg} is that the
dependence on deformation parameters can be extracted directly from the
single-variable function $\ell$. For any deformation parameter
$\alpha$ that enters $\ell$, one finds
\begin{equation}
\left.
\frac{\partial\mathcal L}{\partial\alpha}
\right|_{U,V}
=
\left.
\frac{\partial\ell}{\partial\alpha}
\right|_{\tau},
\qquad
\alpha\in\{\lambda,\gamma\}.
\label{id}
\end{equation}
This identity follows from the implicit definition of $\tau$ in
\eqref{Lg} and will allow us to derive the deformation flows without
explicitly solving for $\tau(U,V)$.

\paragraph{Lemma 1.}
$\partial_\lambda\mathcal L\big|_{U,V,\gamma}=\partial_\lambda\ell\big|_{\tau,\gamma}$.

\emph{Proof.} Write $h=\dot\ell$. At fixed $U,V$ the constraint in \eqref{Lg} gives $(1+2Uh_\tau/h^3)\,\delta\tau=-2U h_\lambda\,\delta\lambda/h^3$. Varying \eqref{Lg},
$\delta\mathcal L=\ell_\lambda\delta\lambda+h(1+2Uh_\tau/h^3)\delta\tau+2Uh_\lambda\delta\lambda/h^2=\ell_\lambda\delta\lambda$. $\square$
The same argument applies verbatim to any other parameter entering $\ell$ (such as $\gamma$ or the coupling $\mu$ used below).

\subsection{One-scale Courant--Hilbert solutions}

Consider a self-dual theory with one dimensionful coupling $\lambda$ and a
dimensionless parameter $\gamma$. In four dimensions,  we consider the associated CH-function, expanded as follows:
\begin{equation}\label{Lexpan}
	\ell (\tau) =e^\gamma \,  \tau  +\sum_{n=1}^\infty c_n \, \lambda^n\, e^{(n+1)\gamma } \, \tau^{n+1}\,,
\end{equation}
Here, $[\ell]=[\tau]=4,$ and $[\lambda]=-4$ and it is therefore natural to introduce the dimensionless combination
$z=e^\gamma\lambda\tau$.
The most general one-scale Courant--Hilbert function may then be written as
\begin{equation}
\ell(\tau;\lambda,\gamma)
=
\frac{1}{\lambda}f(z).
\label{lCH}
\end{equation}
No assumption about the analyticity of $f(z)$ is needed for the following
derivation. The weak-field normalization corresponding to the Maxwell
limit requires
$f(0)=0$, and $f'(0)=1$.

We introduce
\begin{equation}
a(z)=zf'(z),
\qquad
b(z)=f(z)-zf'(z)=f(z)-a(z).
\label{ab2}
\end{equation}
The two combinations of $\ell$ that enter the stress tensor then take the
particularly simple form
\begin{equation}
\tau\dot\ell=\frac{a(z)}{\lambda},
\qquad
\ell-\tau\dot\ell=\frac{b(z)}{\lambda}.
\label{ab}
\end{equation}

For a general Courant--Hilbert solution, the energy--momentum tensor can be
expressed in terms of the Maxwell tensor as
\begin{equation}
T_{\mu\nu}
=
\frac{\tau\dot\ell}{U+V}\,
T_{\mu\nu}^{\rm Max}
+
\left(\ell-\tau\dot\ell\right)g_{\mu\nu}.
\label{Tdecomp}
\end{equation}
Consequently,
\begin{equation}
T_{\mu\nu}T^{\mu\nu}
=
4\left[
(\tau\dot\ell)^2+
(\ell-\tau\dot\ell)^2
\right],
\label{TT}
\end{equation}
while its trace is
\begin{equation}
T^\mu{}_\mu
=
4\left(\ell-\tau\dot\ell\right).
\label{Ttrace}
\end{equation}
It is therefore sufficient to consider the two quadratic stress-tensor
invariants
\begin{equation}
X\equiv T_{\mu\nu}T^{\mu\nu}
=
\frac{4\left[a(z)^2+b(z)^2\right]}{\lambda^2},
\label{X}
\end{equation}
and
\begin{equation}
Y\equiv\left(T^\mu{}_\mu\right)^2
=
\frac{16b(z)^2}{\lambda^2}.
\label{Y}
\end{equation}
Their ratio is independent of the deformation scale,
\begin{equation}
\frac{Y}{X}
=
\frac{4b(z)^2}{a(z)^2+b(z)^2}.
\label{YoverX}
\end{equation}
This dimensionless ratio will play the role of the invariant coordinate
that allows the auxiliary parameter $z$ to be eliminated.

\subsection{Parametric form of the exact flows}

Using \eqref{id} and \eqref{lCH}, the derivatives of the Lagrangian with
respect to the two deformation parameters are
\begin{equation}
\frac{\partial\mathcal L}{\partial\lambda}
=
-\frac{b(z)}{\lambda^2},
\qquad
\frac{\partial\mathcal L}{\partial\gamma}
=
\frac{a(z)}{\lambda}.
\label{flows_ab}
\end{equation}
The marginal flow therefore takes the universal form
\begin{equation}
\boxed{
\frac{\partial\mathcal L}{\partial\gamma}
=
\tau\dot\ell(\tau).
}
\label{gamma_flow}
\end{equation}
This reproduces the universal root-$T\bar T$ flow obtained previously
\cite{Babaei-Aghbolagh:2025cni}.

The $\lambda$-flow can first be expressed directly through the trace,
\begin{equation}
\frac{\partial\mathcal L}{\partial\lambda}
=
-\frac{1}{4\lambda}\,T^\mu{}_\mu.
\label{trace_flow}
\end{equation}
Although exact, this form is not autonomous since it contains an explicit
factor of $\lambda$. A genuinely invariant representation follows from
\eqref{X} and \eqref{flows_ab}:
\begin{equation}
\boxed{
\frac{\partial\mathcal L}{\partial\lambda}
=
X\,F(z),
\qquad
F(z)
=
-\frac{b(z)}
{4\left[a(z)^2+b(z)^2\right]}.
}
\label{f2}
\end{equation}
Equation \eqref{f2} is an exact, non-perturbative parametric flow valid for
every one-scale self-dual nonlinear electrodynamics theory described by
\eqref{lCH}. Notice that the deformation parameter appears only through
the auxiliary parameter $z$. In particular,
\begin{equation}
F(0)=\frac{f''(0)}{8}.
\end{equation}

The dimensions are consistent with the fact that in four dimensions
$[\lambda]=-4$ and $[\mathcal L]=4$, so that
\begin{equation}
\left[\frac{\partial\mathcal L}{\partial\lambda}\right]
=
8,
\qquad
[X]=[Y]=8.
\end{equation}

\subsection{Autonomous stress-tensor  exact $T\bar{T}$-flow}

The parametric representation \eqref{f2} can be converted into a flow
involving only stress-tensor invariants whenever the map
\begin{equation}
z\longmapsto
\mathcal R(z)\equiv\frac{Y}{X}
\label{Rdef}
\end{equation}
is one-to-one on the physical branch under consideration. In that case,
one may invert \eqref{YoverX} and write
\begin{equation}
z=z(\mathcal R).
\end{equation}
Defining
\begin{equation}
\mathcal F(\mathcal R)
\equiv
F\bigl(z(\mathcal R)\bigr),
\end{equation}
the exact flow \eqref{f2} becomes
\begin{equation}
\boxed{
\frac{\partial\mathcal L}{\partial\lambda}
=
X\,\mathcal F\left(\frac{Y}{X}\right).
}
\label{autonomous_flow}
\end{equation}
This is the central result: the irrelevant deformation of any one-scale
self-dual NED theory can be written as an autonomous function of the two
quadratic stress-tensor invariants. In particular, the right-hand side of
\eqref{autonomous_flow} contains no explicit dependence on
$\lambda$, $\tau$, $U$, or $V$.

The branch structure can be made more transparent by introducing the
signed ratio
\begin{equation}
r(z)
\equiv
-\frac{b(z)}{a(z)}.
\label{rdef}
\end{equation}
On a physical branch satisfying
$\tau\dot\ell=a(z)/\lambda>0$, this ratio can be expressed entirely in
terms of stress-tensor invariants,
\begin{equation}
r
=
-\frac{T^\mu{}_\mu}{\sqrt{4X-Y}}.
\label{rinvariant}
\end{equation}
Moreover,
\begin{equation}
\frac{Y}{X}
=
\frac{4r^2}{1+r^2}.
\label{Yr}
\end{equation}
Introducing the principal non-negative variable
\begin{equation}
\rho\equiv\sqrt{\frac{Y}{X}},
\label{rho}
\end{equation}
we obtain
\begin{equation}
r
=
-\operatorname{sgn}(T^\mu{}_\mu)
\frac{\rho}{\sqrt{4-\rho^2}}.
\label{r_rho}
\end{equation}
The flow function in \eqref{f2} may then be written as
\begin{equation}
\boxed{
F(z)
=
\frac{r(z)}
{4a(z)\left[1+r(z)^2\right]}.
}
\label{Fr}
\end{equation}
Thus, whenever $r(z)$ is invertible on the chosen branch, the auxiliary
variable can equivalently be eliminated through its inverse $z=z(r)$.
If the map develops turning points, the autonomous flow is defined
branchwise on the corresponding intervals of monotonicity.

\subsection{Operator expansion}

The invariant form of the flow also provides a direct characterization of
its operator content. Expanding the flow function around the $\rho \to 0$,
\begin{equation}
F(\rho)
=
\sum_{m=0}^{\infty}C_m\rho^m,
\label{Fexp}
\end{equation}
and using $\rho=\sqrt{Y/X}$, we find
\begin{equation}
\frac{\partial\mathcal L}{\partial\lambda}
=
\sum_{m=0}^{\infty}
C_m\,
Y^{m/2}X^{1-m/2}.
\label{flow_expansion}
\end{equation}
The even-$m$ terms involve integer powers of the quadratic stress-tensor
invariants, whereas the odd-$m$ terms generate half-integer powers.
Hence the flow is built from integer powers of $X$ and $Y$ if and only if
all odd coefficients $C_{2k+1}$ vanish. This is a statement about the
operator content of the flow only. In the examples below we will see that
it distinguishes the generalized Born--Infeld theory from the other
theories considered.

\subsection{Causality condition and stress tensor}
\label{sec:causality}

Let $R=\sqrt{S^2+P^2}$, so that
\begin{equation}
  U+V=R,\qquad V-U=S,\qquad 4UV=P^2 .
\end{equation}
The Courant--Hilbert function is normalized by $\ell(0)=0$. Causality and convexity of this class of theories are equivalent to~\cite{Russo:2024llm,Russo:2024xnh}
\begin{equation}
  \dot\ell\ge1,\qquad \ddot\ell\ge0 .
  \label{eq:causal}
\end{equation}
In a frame with $\mathbf E\parallel\mathbf B$ one has $V=E^2/2$ and $U=B^2/2$, so $\tau=\tfrac12E^2+\tfrac12B^2/\dot\ell^{\,2}$. Purely magnetic configurations have $V=0$; purely electric ones have $U=0$ and $\tau=V$.


The stress tensor of the Courant--Hilbert Lagrangian \eqref{Lg} is given by \eqref{Tdecomp}, where $T^{\rm Max}_{\mu\nu}$ is traceless and satisfies $T^{\rm Max}_{\mu\rho}T^{{\rm Max}\,\rho}{}_{\nu}=R^2g_{\mu\nu}$. In the physical sections below it is convenient to work with the unscaled combinations
\begin{equation}
  a\equiv\tau\dot\ell,\qquad b\equiv\ell-\tau\dot\ell ,
\end{equation}
which in the one-scale notation of \eqref{ab} are $a(z)/\lambda$ and $b(z)/\lambda$. The trace follows from $T^\mu{}_\mu=4(\mathcal L-S\mathcal L_S-P\mathcal L_P)$ and $S\mathcal L_S+P\mathcal L_P=V\mathcal L_V+U\mathcal L_U$ (which uses $P^2=4UV$), and reproduces \eqref{Ttrace}--\eqref{Y}:
\begin{equation}
  T^\mu{}_\mu=4b,\qquad
  X=T_{\mu\nu}T^{\mu\nu}=4(a^2+b^2),\qquad Y=(T^\mu{}_\mu)^2=16\,b^2 .
\end{equation}

The traceless part $\hat T_{\mu\nu}=T_{\mu\nu}-\tfrac14g_{\mu\nu}T^\rho{}_\rho$ has $\hat T_{\mu\nu}\hat T^{\mu\nu}=4a^2$. The eigenvalues of $T^\mu{}_\nu$ are $a+b=\ell$ (twice) and $b-a=-(2\tau\dot\ell-\ell)$ (twice), so we define
\begin{equation}
  \mathcal E\equiv2\tau\dot\ell-\ell,\qquad p\equiv\ell ,
\end{equation}
the energy density and the transverse pressure in the frame $\mathbf E\parallel\mathbf B$. The variable used throughout the analysis of the singular points is
\begin{equation}
  \kappa=\frac{|\ell-\tau\dot\ell|}{|\tau\dot\ell|}
  =\frac{|T^\mu{}_\mu|}{2\sqrt{\hat T_{\mu\nu}\hat T^{\mu\nu}}}
  =\frac{\mathcal E-p}{\mathcal E+p},
  \qquad
  \frac p{\mathcal E}=\frac{1-\kappa}{1+\kappa}.
  \label{eq:kappaphys}
\end{equation}
It is the ratio of the scale-breaking part of $T_{\mu\nu}$ (its trace) to the scale-covariant part $\hat T_{\mu\nu}$.

Under \eqref{eq:causal} and $\ell(0)=0$, convexity gives $0=\ell(0)\ge\ell(\tau)-\tau\dot\ell(\tau)$, i.e.\ $b\le0$, and $\ell>0$ for $\tau>0$. Hence
\begin{equation}
  \mathcal E\ge p>0,\qquad T^\mu{}_\mu\le0,\qquad \kappa=1-\frac{\ell}{\tau\dot\ell}\in[0,1).
\end{equation}
The dominant energy condition $\mathcal E\ge|p|$ is equivalent to $b\le0$, so it holds automatically in the causal sector. Here and below the energy density is denoted by $\mathcal E$, to distinguish it from the ratio $\rho=\sqrt{Y/X}$ of \eqref{rho} that labels the flow functions.

\section{Generalized Born--Infeld theory}
\label{sec:GBI_flow}

\subsection{Exact flow}

The self-energy of a charged particle diverges in Maxwell's theory. Born-Infeld theory, one of the well-known nonlinear electrodynamics theories, was introduced in~\cite{Born:1934gh} to render this self-energy finite.
The Courant--Hilbert function defining the generalized Born--Infeld (BI) theory is~\cite{Russo:2024llm}
\begin{equation}\label{CHBI}
\ell_{\mathrm{GBI}}(\tau;\lambda,\gamma)
=
\frac{1}{\lambda}
\left(
1-\sqrt{1-2e^\gamma\lambda\tau}
\right).
\end{equation}
Solving the second equation in~\eqref{Lg} for the Courant-Hilbert function~\eqref{CHBI} and substituting into the first equation in~\eqref{Lg}, the  generalized Born--Infeld Lagrangian of the theory is found to be
\begin{equation}
	\label{MMB}
	\quad  \mathcal{L}_{GBI} =  \frac{1}{\lambda} \Big[1- \sqrt{\left(1+ 2 \, \lambda e^{-\gamma} \,U\right)\left(1 -2 \, \lambda e^\gamma \,V\right)} \Big] \,. 
\end{equation}
Born-Infeld theory in~\eqref{MMB} is a causal, duality-invariant theory that satisfies the partial differential equation~\eqref{CH}.

The Courant--Hilbert function in~\eqref{CHBI} can then be written in the  one-scale variable $z=e^\gamma\lambda\tau$,
\begin{equation}\label{fBI}
\ell_{\mathrm{GBI}}(\tau;\lambda)=\frac{1}{\lambda}f(z), \qquad f(z)=1-\sqrt{1-2z}.
\end{equation}
Substituting~\eqref{fBI} into~\eqref{ab2}, the generalized Born-Infeld theorem gives the two functions $a(z)$ and $b(z)$ as
\begin{equation}
a(z)=\frac{z}{\sqrt{1-2z}},
\qquad
b(z)=-\frac{(1-\sqrt{1-2z})^{2}}{2\sqrt{1-2z}}.
\end{equation}
Given $a(z)$ and $b(z)$ for generalized Born-Infeld theory, the dimensionless function $F_{BI}(z)$ is readily obtained as
\begin{equation}\label{fbii}
F_{GBI}(z)=\frac{\sqrt{1-2z}}{8(1-z)}.
\end{equation}
From the ratio $r=-\,b(z)/a(z)$ we obtain $r$ in terms of $z$, and hence $z(r)$ as a function of $r$:
\begin{equation}\label{zbi}
r(z)=\frac{1-\sqrt{1-2z}}{1+\sqrt{1-2z}},
\quad \Longrightarrow\quad  z(r)=\frac{2 r}{(1+r)^2}.
\end{equation}
Substituting $z(r)$ from the second equation in~\eqref{zbi} into~\eqref{fbii}, the two-dimensional function $F_{\mathrm{BI}}(r)$ is obtained as a function of $r$:\begin{equation}
F_{\mathrm{GBI}}(r)=\frac{1-r^2}{8(1+r^2)}. 
\end{equation}
Using~\eqref{autonomous_flow} and~\eqref{r_rho} we obtain
\begin{equation}\label{ttbi}
\frac{\partial\mathcal L_{\mathrm{GBI}}}{\partial\lambda}
=X \left(\frac18-\frac{\rho^2}{16}\right) = 
\frac18 \left(T_{\mu\nu}T^{\mu\nu}
-\frac12 T_{\mu}{}^{\mu}T_{\nu}{}^{\nu}\right). 
\end{equation}
Thus, although the generalized Born-Infeld (GBI) Lagrangian is a nonlinear function of the electromagnetic invariants, its exact irrelevant flow is quadratic in the stress tensor. This flow was previously obtained in Ref.~\cite{Babaei-Aghbolagh:2022uij}, and we reproduce it here using the autonomous stress-tensor  exact $T\bar{T}$-flow approach, demonstrating that this approach works well.

We present an energy analysis based on Sec.~\eqref{sec:causality}. This allows us to examine how the energy of the system behaves under the deformation and to identify which branches of the solution are physically admissible.
\begin{equation}
X=4\big[(\tau\dot\ell)^{2}+(\ell-\tau\dot\ell)^{2}\big],\qquad
Y=16\,(\ell-\tau\dot\ell)^{2},
\end{equation}
and $a=\tau\dot\ell$, $b=\ell-\tau\dot\ell$, one finds $2X-Y=8(a^{2}-b^{2})$. Using in addition the eigenvalues $a+b=\ell=p$ and $a-b=2\tau\dot\ell-\ell=\mathcal E$ of $T^\mu{}_\nu$ introduced in section~\ref{sec:causality}, the flow \eqref{ttbi} becomes
\begin{equation}
   \frac{\partial\mathcal L_{GBI}}{\partial\lambda}=\tfrac18\Big(T_{\mu\nu}T^{\mu\nu}-\tfrac12(T^\mu{}_\mu)^2\Big)=\tfrac12(a^2-b^2)=\tfrac12\,\mathcal E\,p
  =\tau\ell\dot\ell-\tfrac12\ell^2 ,
  \label{eq:ZZphys}
\end{equation}
the four-dimensional $T\bar T$-like operator. Thus $\frac{\partial\mathcal L_{GBI}}{\partial\lambda}$ is a polynomial in $(\ell,\tau\dot\ell)$ and has no singularities. The GBI theory is the $q=\tfrac12$ member of the family studied in section~\ref{sec:q_deformed}, and it is the only member whose flow is polynomial in the stress tensor.

\subsection{Physical domain }

The physical domain of the GBI solution is determined by the reality
condition in \eqref{CHBI}. For $\lambda>0$ and $\gamma\geq0$,
\begin{equation}
0\leq\tau\leq
\frac{1}{2e^\gamma\lambda},
\qquad
0\leq z\leq\frac12,
\qquad
0\leq s\leq1.
\label{GBI_domain}
\end{equation}
The corresponding range of the signed ratio $r=-b/a$ follows directly
from \eqref{zbi}:
\begin{equation}
0\leq r\leq1.
\label{GBI_r_domain}
\end{equation}
The endpoints are unambiguous.

The $F_{\mathrm{GBI}}(\rho)$ is exact already at quadratic
order,
\begin{equation}
F_{\mathrm{GBI}}(\rho)
=
\frac18-\frac{\rho^2}{16},
\end{equation}
so that
\begin{equation}
C_0=\frac18,
\qquad
C_1=0,
\qquad
C_2=-\frac1{16},
\qquad
C_{m\geq3}=0.
\end{equation}
The absence of odd powers of $\rho$ is therefore manifest: the flow is
built from integer powers of $X$ and $Y$ only. The GBI model consequently realizes the simplest possible
example of the general autonomous flow: the full nonlinear deformation
resums exactly into the quadratic stress-tensor operator in
\eqref{ttbi}.


\section{The $q$-deformed theory}
\label{sec:q_deformed}

\subsection{Exact autonomous flow}

We begin with the one-parameter family of Courant--Hilbert functions
\begin{equation}
\ell_q(\tau;\lambda,\gamma)
=
\frac{1}{\lambda}
\left[
1-
\left(
1-\frac{e^\gamma\lambda\tau}{q}
\right)^q
\right],
\label{eq:ellq}
\end{equation}
which reduces to the ModMax function $\ell_1=e^\gamma\tau$ at $q=1$
and to the Maxwell seed in the limit $\lambda\to0$. It is convenient
to introduce the dimensionless variables
\begin{equation}
z=e^\gamma\lambda\tau,
\qquad
s=1-\frac{z}{q},
\end{equation}
in terms of which the CH function takes the one-scale form
\begin{equation}
\ell_q=\frac{1}{\lambda}f_q(z),
\qquad
f_q(s)=1-s^q,
\end{equation}
with real branch $s>0$. The first two derivatives are
\begin{equation}
\dot{\ell}=e^\gamma s^{q-1},
\qquad
\ddot{\ell}
=
e^{2\gamma}\lambda
\frac{1-q}{q}s^{q-2}.
\end{equation}
Imposing the causality conditions $\dot{\ell}\geq1$ and
$\ddot{\ell}\geq0$, and assuming $\gamma\geq0$ and $q>0$, selects the
physical branches
\begin{equation}
\begin{array}{c|c|c|c}
\text{Range of }q & \text{Sign of }\lambda
& \text{Range of }s & \text{Range of }\tau \\ \hline
0<q<1 & \lambda>0
& 0<s\leq1
& 0\leq\tau<q e^{-\gamma}/\lambda
\\[1mm]
q>1 & \lambda<0
& 1\leq s<\infty
& 0\leq\tau<\infty
\end{array}.
\end{equation}
In particular, the $q>1$ theories, including $q=3/2$, are causal only
on the negative-$\lambda$ branch.

For this family the two functions \eqref{ab2} read
\begin{equation}
a_q(s)=q(1-s)s^{q-1},
\qquad
b_q(s)=1-qs^{q-1}+(q-1)s^q,
\end{equation}
so that the signed invariant ratio \eqref{rdef} becomes
\begin{equation}\label{rq}
r_q(s)
=
\frac{q+(1-q)s-s^{1-q}}
     {q(1-s)}.
\end{equation}
At the vacuum point $s=1$ this expression takes the indeterminate form
$0/0$. The singularity is nevertheless removable: expanding the
numerator and denominator around $s=1$ and applying L'H\^opital's rule
gives
\begin{equation}
\lim_{s\to1}r_q(s)=0,
\end{equation}
so that $r_q(s)$ admits a unique continuous extension to the vacuum,
\begin{equation}
r_q(1)\equiv0.
\end{equation}
On each physical branch $r_q(s)$ is continuous and strictly monotonic,
hence one-to-one, and therefore globally invertible. The inverse is
single-valued and is fixed by
\begin{equation}
s_q^{\,1-q}
=
q+(1-q)s-q(1-s)r.
\label{eq:sqdef}
\end{equation}
After clearing the denominator in \eqref{rq}, $s=1$ reappears as a
formal root for every $r$; for $r\neq0$ this is a spurious $0/0$ root.
The physical solution is selected by the branch conditions
\begin{equation}
0<s_q(r)<1
\quad (0<q<1),
\qquad
s_q(r)>1
\quad (q>1),
\end{equation}
with ranges
\begin{equation}
0< r<1
\quad (0<q<1),
\qquad
0< r<\frac{q-1}{q}
\quad (q>1).
\end{equation}
The exact flow function in terms of the signed invariant $r$ is
\begin{equation}\label{Fq}
F_q(r)
=
\frac{r\,s_q^{1-q}}
{4q\left(1-s_q\right)(1+r^2)},
\end{equation}
and, using the relation \eqref{r_rho} between $r$ and the principal
variable $\rho$,
\begin{equation}
\rho=\frac{2r}{\sqrt{1+r^2}},
\qquad
r=\frac{\rho}{\sqrt{4-\rho^2}},
\end{equation}
the exact irrelevant flow takes the autonomous form
\begin{equation}\label{fq}
\frac{\partial\mathcal L_q}{\partial\lambda}
=
X\,\mathcal{ F}_q\!\left(\sqrt{\frac{Y}{X}}\right),
\end{equation}
with
\begin{equation}
\mathcal{ F}_q(\rho)
=
\frac{\rho\sqrt{4-\rho^2}}{16q}
\frac{s_q(\rho)^{\,1-q}}
     {1-s_q(\rho)}.
\end{equation}
Here $s_q(\rho)$ denotes the solution of
\begin{equation}
s_q^{\,1-q}
=
q\left(
1-\frac{\rho}{\sqrt{4-\rho^2}}
\right)
+
\left(
1-q+\frac{q\rho}{\sqrt{4-\rho^2}}
\right)s_q.
\end{equation}
Since $r\mapsto s_q(r)$ is strictly monotonic on the physical branch,
its inverse is single-valued, and $\mathcal{F}_q(\rho)$ is uniquely
determined by the choice of branch. No explicit dependence on
$\lambda$, $\gamma$, $\tau$, $U$, or $V$ survives in \eqref{fq}: the
flow is exact, nonperturbative, and autonomous. The physical domains are
\begin{equation}
0<\rho<\sqrt{2}
\quad (0<q<1),
\qquad
0<\rho<
\frac{2(q-1)}
{\sqrt{q^2+(q-1)^2}}
\quad (q>1).
\end{equation}

Expanding $\mathcal{F}_q(\rho)$ around the weak-field point $\rho=0$
gives $F_q(\rho)=\sum_{m\geq0}C_m(q)\rho^m$, with
\begin{align}\label{eFq}
F_q(\rho)
={}&
\frac{1-q}{8q}
+\frac{2q-1}{12q}\,\rho
-\frac{(q+1)(8q-7)}
{288q(q-1)}\,\rho^2
\nonumber\\[1mm]
&-\frac{(2q-1)
\left(43q^2-88q+49\right)}
{4320q(q-1)^2}\,\rho^3
\nonumber\\[1mm]
&+\frac{(2q-1)^2(q-2)(q+1)}
{12960q(q-1)^3}\,\rho^4
\nonumber\\[1mm]
&-\frac{(2q-1)
\left(
2335q^4-10088q^3+16926q^2
-12512q+3499
\right)}
{4354560q(q-1)^4}\,\rho^5
\nonumber\\[1mm]
&+\frac{(2q-1)^2(q-2)(q+1)
\left(331q^2-541q+268\right)}
{10886400q(q-1)^5}\,\rho^6
+O\!\left(\rho^7\right).
\end{align}
These coefficients agree exactly with those previously obtained in
\cite{Babaei-Aghbolagh:2026vkm}. Equivalently, the flow can be written
as a series in the two stress-tensor invariants,
\begin{equation}
\frac{\partial\mathcal L_q}{\partial\lambda}
=
\sum_{m=0}^{\infty}
C_m(q)\,
Y^{m/2}X^{1-m/2}.
\end{equation}

\subsection{The case $q=1/2$}
For $q=\frac{1}{2}$ the inversion is exact. Setting $u=\sqrt{s}$,
equation \eqref{rq} becomes
\begin{equation}
r=\frac{1-u}{1+u},
\end{equation}
which inverts to
\begin{equation}
u=\frac{1-r}{1+r},
\qquad\Longrightarrow\qquad
s(r)=
\left(
\frac{1-r}{1+r}
\right)^2.
\end{equation}
Substituting into \eqref{Fq} gives
\begin{equation}
F_{1/2}(r)
=
\frac{1-r^2}{8(1+r^2)},
\qquad\Longrightarrow\qquad
F_{1/2}(\rho)
=
\frac{1}{8}-\frac{\rho^2}{16}.
\end{equation}
Comparison with the general expansion \eqref{eFq} yields
\begin{equation}
C_0\!\left(\tfrac{1}{2}\right)=\frac{1}{8},
\qquad
C_2\!\left(\tfrac{1}{2}\right)=-\frac{1}{16},
\qquad
C_m\!\left(\tfrac{1}{2}\right)=0
\quad (m\neq0,2).
\end{equation}
The infinite stress-tensor expansion therefore truncates exactly at
second order, and the exact irrelevant flow reduces to
\begin{equation}
\frac{\partial\mathcal L_{1/2}}{\partial\lambda}
=
\frac{1}{8}X-\frac{1}{16}Y,
\end{equation}
the linear $T\bar T$-like deformation that generates the generalized
Born--Infeld theory.

\subsection{The case $q=3/4$}
For $q=\frac{3}{4}$ it is convenient to set $u=s^{1/4}$. Equation
\eqref{rq} then becomes
\begin{equation}
r
=
\frac{3+u^{4}-4u}{3(1-u^{4})}.
\end{equation}
After removing the spurious root $u=1$, the physical inverse satisfies
\begin{equation}
(1+3r)(u^{3}+u^{2}+u)
=
3(1-r),
\end{equation}
with a unique physical root $u\in(0,1)$. From \eqref{Fq},
\begin{equation}
\mathcal{F}_{3/4}(r)
=
\frac{r u}
{3(1-u^{4})(1+r^{2})},
\end{equation}
whose expansion in the principal variable $\rho$ reads
\begin{equation}
\begin{aligned}
\mathcal{F}_{3/4}(\rho)
={}&
\frac{1}{24}
+\frac{\rho}{18}
-\frac{7\rho^{2}}{216}
-\frac{23\rho^{3}}{1296}
\\
&+
\frac{7\rho^{4}}{1944}
-\frac{869\rho^{5}}{186624}
+\frac{155\rho^{6}}{46656}
+O(\rho^{7}).
\end{aligned}
\end{equation}
This reproduces exactly the coefficients previously obtained for the
$q=\frac{3}{4}$ model in \cite{Babaei-Aghbolagh:2026vkm}. The odd
coefficients are non-zero, so, in contrast with $q=\frac12$, the flow
contains half-integer powers of $Y/X$.

\subsection{The case $q=3/2$}
The case $q=3/2$ belongs to the $q>1$ branch, which is causal only for
$\lambda<0$. Its flow follows from the general formulas above with
$q=3/2$, the physical root $s_q(r)$ being selected by $s_q(r)>1$. The
resulting autonomous flow again takes the form
$\partial_\lambda\mathcal{L}_{3/2}=X\mathcal{F}_{3/2}(Y/X)$, with
$\mathcal{F}_{3/2}$ single-valued on the physical branch and
containing half-integer powers of $Y/X$, since the coefficient
$C_1=(2q-1)/(12q)$ in \eqref{eFq} does not vanish.
This is the theory treated in closed form in section~\ref{sec:NM}: with the coupling $\lambda'=-\tfrac32\lambda<0$ the member $q=\tfrac32$ coincides with the no-$\tau$-maximum Courant--Hilbert function.

\subsection{Causality, quartic coupling and the flow operator}
\label{sec:q_phys}

The family \eqref{eq:ellq} was introduced in section~\ref{sec:q_deformed}. We write $ s=1-e^\gamma\lambda\tau/q$ for its auxiliary variable, so that $\dot\ell_q=e^\gamma y^{\,q-1}$ and $\ddot\ell_q=\frac{1-q}{q}\,\lambda e^{2\gamma}y^{\,q-2}$, and the expansion near the vacuum reads
\begin{equation}
  \ell_q=e^\gamma\tau+c_q\lambda e^{2\gamma}\tau^2+O(\tau^3),\qquad c_q=\frac{1-q}{2q}.
\end{equation}
For $y\in(0,1]$ ($q<1$, $\lambda>0$) and for $y\ge1$ ($q>1$, $\lambda<0$) one has $y^{q-1}\ge1$ when $\gamma\ge0$, so \eqref{eq:causal} reduces to
\begin{equation}
  \boxed{\;\gamma\ge0,\qquad \frac{\lambda(1-q)}{q}\ge0\;}
  \label{eq:cond}
\end{equation}
i.e.\ $\lambda>0$ for $0<q<1$ and $\lambda<0$ for $q>1$. The parameter $\gamma$ is the ModMax parameter: $\ell=e^\gamma\tau$ gives $\mathcal L=e^\gamma V-e^{-\gamma}U$. The members of the family are:
\begin{itemize}
\item $q=\tfrac12$: $\ell=\lambda^{-1}\big[1-\sqrt{1-2e^\gamma\lambda\tau}\big]$, Born--Infeld--ModMax ($\lambda=1/T$ at $\gamma=0$);
\item $q\to1$: $\ell=e^\gamma\tau$, ModMax; $\lambda$ drops out;
\item $q\to0$ with $\lambda=q\mu$ fixed: logarithmic electrodynamics (section~\ref{sec:logarithmic});
\item $q\to\infty$: $\ell=\lambda^{-1}\big(1-e^{-e^\gamma\lambda\tau}\big)$, exponential electrodynamics ($\lambda<0$);
\item $q=2$: $\ell=e^\gamma\tau+\tfrac14e^{2\gamma}|\lambda|\tau^2$.
\end{itemize}
For $\gamma=0$ the expansion of \eqref{Lg} gives
\begin{equation}
  \mathcal L=S+c_q\lambda\,(S^2+P^2)+O(F^6),
  \label{eq:quartic}
\end{equation}
so \eqref{eq:cond} states that the quartic coefficient $c_q\lambda$ is non-negative, as required by positivity/causality. It vanishes for $q=1$.

\paragraph{Lemma 2.}
For \eqref{eq:ellq} the non-trivial root of \eqref{eq:sqdef} is exactly $s_q=y$, and
\begin{equation}
  \boxed{\; \frac{\partial\mathcal L_{q}}{\partial\lambda}=\partial_\lambda\ell_q\big|_{\tau,\gamma}=\partial_\lambda\mathcal L
  =\frac{\tau\dot\ell-\ell}{\lambda}=-\frac{T^\mu{}_\mu}{4\lambda}
  =\frac{1}{\lambda^2}\Big[q(1-y)\,y^{q-1}+y^q-1\Big]\;}
  \label{eq:ZZ2phys}
\end{equation}

\emph{Proof.} From \eqref{eq:ellq}, $\ell=(1-y^q)/\lambda$ and $a=\tau\dot\ell=q(1-y)y^{q-1}/\lambda$, so $\kappa=1-\ell/a$. Substituting $s=y$ into \eqref{eq:sqdef} gives the identity $y^{1-q}=y^{1-q}$. Then $\lambda=q(1-s_q)/(e^\gamma\tau)$ and $e^\gamma\tau=a\,s_q^{1-q}$ turn \eqref{Fq} into $(a-\ell)/\lambda$. Finally $\partial_\lambda\ell_q|_\tau=\lambda^{-2}\big[y^q-1+q(1-y)y^{q-1}\big]$ agrees with it. $\square$

\paragraph{Consequences.}
\begin{enumerate}
\item \eqref{eq:ZZ2phys} is the identity for the dimensionful coupling: $\mathcal L=\lambda^{-1}\Phi(\lambda S,\lambda P)$ at fixed $\gamma$ implies $T^\mu{}_\mu=-4\lambda\,\partial_\lambda\mathcal L$. What depends on $q$ is how $T^\mu{}_\mu/\lambda$ is expressed through the invariants of $T_{\mu\nu}$ alone: $ \frac{\partial\mathcal L_{q}}{\partial\lambda}=X\mathcal F_q(\sqrt{Y/X})$, with $\lambda$ eliminated through $s_q$.
\item $ \frac{\partial\mathcal L_{GBI}}{\partial\lambda}= \frac{\partial\mathcal L_{q}}{\partial\lambda}|_{q=1/2}$ holds exactly for every $\gamma$: the $T\bar T$ operator generates the Born--Infeld--ModMax flow. It is the only member of the family that is polynomial in $T_{\mu\nu}T^{\mu\nu}$ and $(T^\mu{}_\mu)^2$.
\item The limit $q\to0$ (with $\lambda=q\mu$) gives the operator $ \frac{\partial\mathcal L_{\log}}{\partial\lambda}$ of logarithmic electrodynamics, treated in section~\ref{sec:logarithmic}. The limit $q\to\infty$ gives the exponential model, which is a different operator.
\item The operator $ \frac{\partial\mathcal L_{q}}{\partial\lambda}$ is defined exactly on the image of the family: a triple $(\ell,\tau,\dot\ell)$ gives a real $s_q$ if and only if it is realised by some $\ell_q$.
\end{enumerate}

\subsection{Physical meaning of the singular points}
\label{sec:q_sing}

All singular points of the flow lie on the one-dimensional family $\kappa\in[0,1)$ and are interpreted below in the field variable $\tau$.

\paragraph{(i) $\kappa\to0$: the conformal point.}
Since $\kappa=1-\ell/(\tau\dot\ell)\simeq c_q\lambda e^\gamma\tau$, this limit is $\tau\to0$, i.e.\ $U,V\to0$, or equivalently $\lambda\to0$. There $T^\mu{}_\mu\to0$ and $p\to\mathcal E$, and
\begin{equation}
   \frac{\partial\mathcal L_{q}}{\partial\lambda}\simeq c_q\,e^{2\gamma}\tau^2 ,
\end{equation}
which is the quartic coupling \eqref{eq:quartic}: $ \frac{\partial\mathcal L_{q}}{\partial\lambda}|_{\lambda=0}$ is the $T\bar T$-like operator evaluated on the seed (Maxwell or ModMax) stress tensor. In the variable $\tau$ the operator is analytic here. The odd powers of $\rho\propto|T^\mu{}_\mu|$ in the weak-field expansion \eqref{eFq}, equivalently a cusp $\propto|\ell-\tau\dot\ell|$ in the variable $\kappa$, are an artefact of using $\kappa\ge0$ as a coordinate: $\kappa\propto\tau$ on the physical domain, where $T^\mu{}_\mu\le0$ and $|T^\mu{}_\mu|=-T^\mu{}_\mu$ is analytic. $ \frac{\partial\mathcal L_{q}}{\partial\lambda}$ is a polynomial in the invariants $X,Y$ only for $q=\tfrac12$. For $\gamma\neq0$ the only genuine non-analyticity at $F=0$ is the one inherited from ModMax through $R=\sqrt{S^2+P^2}$.

\paragraph{(ii) $\kappa\to1$ for $q<1$: the critical field.}
Here $y\to0$, $\tau\to\tau_{\max}=qe^{-\gamma}/\lambda$ and $\dot\ell_q=e^\gamma y^{q-1}\to\infty$. Since $\tau=V+U/\dot\ell^2$, the Lagrangian is defined for all $U\ge0$ if and only if
\begin{equation}
  V<V_{\max}=\frac{q\,e^{-\gamma}}{\lambda},
\end{equation}
i.e.\ in the frame $\mathbf E\parallel\mathbf B$ the electric field is bounded, $E^2<2qe^{-\gamma}/\lambda$. For $q=\tfrac12$, $\gamma=0$ this is the Born--Infeld bound $E^2<1/\lambda=T$, while magnetic fields are unbounded. At the bound
\begin{equation}
  \mathcal E\to\infty,\qquad p=\ell\to\frac1\lambda,\qquad \frac p{\mathcal E}\to0^+,\qquad  \frac{\partial\mathcal L_{q}}{\partial\lambda}\simeq\frac{\mathcal E}{2\lambda}\to\infty ,
\end{equation}
while the normalised operator $ \frac{\partial\mathcal L_{q}}{\partial\lambda}/(\tau\dot\ell)^2\simeq1-\kappa$ vanishes. The root behaves as $s_q=y\simeq(q\varepsilon)^{1/(1-q)}$ with $\varepsilon=1-\kappa$. In the physical variable the endpoint is a branch point for every non-integer $q$, because $\ell_q\simeq\lambda^{-1}(1-y^q)$: a square root for $q=\tfrac12$ (Born--Infeld) and a logarithm in the limit $q\to0$. Beyond the critical field the Lagrangian is complex, which signals the limit of validity of the effective description. For integer $1/(1-q)$ (e.g.\ $q=\tfrac12,\tfrac34$) the root $s_q$ is analytic in $\kappa$, but this is only a property of the reparametrisation $\kappa(\tau)$ and does not remove the branch point in $\tau$.

\paragraph{(iii) $\kappa\to\kappa_c=(q-1)/q$ for $q>1$: strong-field asymptotics.}
Here $\lambda<0$ and $y=1+|e^\gamma\lambda|\tau/q$ grows without bound, with $\dot\ell_q\ge e^\gamma$ and $\ddot\ell_q>0$ for all $\tau\ge0$. The theory is causal at all field strengths and has no critical field. The point $\kappa_c$ is reached only for $\tau\to\infty$, where
\begin{equation}
  \ell_q\simeq|\lambda|^{-1}y^q\propto\tau^q,\qquad \frac p{\mathcal E}\to\frac1{2q-1},\qquad  \frac{\partial\mathcal L_{q}}{\partial\lambda}\simeq-\frac{(q-1)\,y^q}{\lambda^2}\to-\infty .
\end{equation}
The root $s_q=y$ diverges there. The non-existence of a real root for $\kappa>\kappa_c$ only says that the family does not realise stress tensors with $p/\mathcal E<1/(2q-1)$. The singularity of $y^q$ at $y=0$ lies at $\tau<0$, outside the physical region, and for integer $q$ the function $\ell_q$ is a polynomial in $\tau$.

\paragraph{(iv) The sign of the flow.}
Since $ \frac{\partial\mathcal L_{q}}{\partial\lambda}=-T^\mu{}_\mu/(4\lambda)$ and $T^\mu{}_\mu\le0$, $ \frac{\partial\mathcal L_{q}}{\partial\lambda}$ has the sign of $\lambda$: positive for $q<1$ and negative for $q>1$. In both sectors increasing $|\lambda|$ raises $\mathcal L$, consistently with \eqref{eq:cond}.

\begin{table}[t]
\centering\small
\begin{tabular}{@{}p{2.1cm}p{3.5cm}p{3.0cm}p{3.0cm}p{3.0cm}@{}}
\hline
Point & Field variables & Stress tensor & Behaviour of $ \frac{\partial\mathcal L_{q}}{\partial\lambda}$ & Interpretation \\
\hline
$\kappa\to0$ & $F\to0$ ($\tau\to0$), $\kappa\simeq c_q\lambda e^\gamma\tau$ & $T^\mu{}_\mu\to0$, $p\to\mathcal E$ & $c_qe^{2\gamma}\tau^2$ & conformal point, quartic coefficient $c_q\lambda\ge0$ \\
$\kappa\to1$, $q<1$ & $\tau\to\tau_{\max}=\frac{q}{\lambda e^{\gamma}}$, $\dot\ell\to\infty$ & $\mathcal E\to\infty$, $p\to1/\lambda$ & $\simeq\mathcal E/2\lambda$ & critical field, branch point $y^q$ \\
$\kappa\to\kappa_c$, $q>1$ & $\tau\to\infty$ & $p/\mathcal E\to\frac1{2q-1}$ & $\simeq-\frac{(q-1)y^q}{\lambda^2}$ & strong-field regime, no critical field \\
$q=\tfrac12$ & Born--Infeld--ModMax & $ \frac{\partial\mathcal L_{GBI}}{\partial\lambda}=\tfrac12\mathcal E p$ & $ \frac{\partial\mathcal L_{q}}{\partial\lambda}= \frac{\partial\mathcal L_{GBI}}{\partial\lambda}$ & only polynomial member \\
$q=1$ & $\ell=e^\gamma\tau$ & $\kappa\equiv0$ & $0$ & ModMax, $\lambda$ drops out \\
\hline
\end{tabular}
\caption{Singular points of $ \frac{\partial\mathcal L_{q}}{\partial\lambda}$ in terms of fields, stress tensor and the causal parameters \eqref{eq:cond}.}
\label{tab:phys-ZZ2}
\end{table}

\section{The no-$\tau$-maximum theory}
\label{sec:NM}

The no-$\tau$-maximum theory provides a flow with half-integer powers of $Y/X$ that can be written in closed form, unlike the generic members of the $q$-deformed family.

\subsection{Courant--Hilbert function and causality}

The theory is defined by the Courant--Hilbert function
\begin{equation}
  \ell_{\rm NM}(\tau;\lambda,\gamma)=\frac{2}{3\lambda}\Big[(1+z)^{3/2}-1\Big],
  \qquad z=e^\gamma\lambda\tau,\quad s=\sqrt{1+z}\ge1,
  \label{eq:ellNM}
\end{equation}
which satisfies $\dot\ell=e^\gamma s\ge1$ and $\ddot\ell=e^{2\gamma}\lambda/(2s)>0$ for all $\tau\ge0$, $\lambda>0$, $\gamma\ge0$, so it is causal. Since $\tau\in[0,\infty)$ there is no critical field. It is the member $q=3/2$ of the family \eqref{eq:ellq} with $\lambda'=-\tfrac32\lambda<0$, as required by \eqref{eq:cond}, and
\begin{equation}
 \frac{\partial\mathcal L_{\mathrm{NM}}}{\partial\lambda}=\partial_\lambda\mathcal L=-\tfrac32\,\frac{\partial\mathcal L_{q}}{\partial\lambda}\big|_{q=3/2,\ \lambda\to\lambda'} .
\end{equation}
In the one-scale form \eqref{lCH} one has
\begin{equation}
f(z)=\frac{2}{3}(s^3-1),
\qquad
f'(z)=s ,
\end{equation}
and the causal branch is $1\leq s<\infty$.

\subsection{Exact autonomous flow}

From \eqref{ab2} one finds
\begin{align}
a(s)
&=s(s-1)(s+1),\\
b(s)
&=-\frac{(s-1)^2(s+2)}{3},
\end{align}
so that \eqref{rdef} gives
\begin{equation}
  r(s)=-\frac ba=\frac13-\frac{2}{3s(s+1)},\qquad
  \frac{dr}{ds}=\frac{2(2s+1)}{3\,s^2(s+1)^2}>0 .
  \label{eq:rNM}
\end{equation}
At the vacuum point $s=1$ both $a$ and $b$ vanish and the ratio
$r=-b/a$ takes the indeterminate form $0/0$. Cancelling the common
factor $s-1$, the expression \eqref{eq:rNM} has the well-defined limit
$\lim_{s\to1^+}r(s)=0$, and we extend the coordinate continuously to the vacuum by setting
$r(1)=0$. Since $dr/ds>0$,
$r(s)$ is strictly monotonic on the entire causal branch, with range
\begin{equation}
0\le r<\frac{1}{3}.
\end{equation}
The strict monotonicity guarantees that the parametric flow defines a
single-valued autonomous function of the stress-tensor invariants. The
inverse is available in closed form,
\begin{equation}
s(r)
=
\frac{1}{2}
\left[
\sqrt{\frac{3(3-r)}{1-3r}}-1
\right],
\end{equation}
so that the auxiliary parameter is eliminated exactly. The flow
function \eqref{Fr} therefore becomes
\begin{equation}
\mathcal{ F}_{\mathrm{NM}}(r)
=
\frac{(1-3r)^2}{96(1+r^2)}
\left[
3+
\sqrt{\frac{3(3-r)}{1-3r}}
\right],
\qquad
r=\frac{\rho}{\sqrt{4-\rho^2}},
\qquad
0\le\rho<\sqrt{\frac25},
\end{equation}
where we used \eqref{r_rho}. Its weak-field expansion is
\begin{equation}
\begin{aligned}
\mathcal{ F}_{\mathrm{NM}}(\rho)
={}&
\frac{1}{16}
-\frac{\rho}{6}
+\frac{25\rho^2}{288}
+\frac{11\rho^3}{432}
+\frac{\rho^4}{324}
\\
&+
\frac{269\rho^5}{62208}
+\frac{23\rho^6}{7776}
+O(\rho^7),
\end{aligned}
\label{eq:FNMexp}
\end{equation}
in agreement with the results of \cite{Babaei-Aghbolagh:2026vkm}. The
exact autonomous stress-tensor flow is therefore
\begin{equation}
\frac{\partial\mathcal L_{\mathrm{NM}}}{\partial\lambda}
=
X\,\mathcal{ F}_{\mathrm{NM}}(r),
\end{equation}
which, like the $q$-deformed flows, contains no explicit dependence on
$\lambda$, $\gamma$, $\tau$, $U$, or $V$. The odd coefficients in the
expansion are non-zero, so the flow contains half-integer powers of
$Y/X$. In terms of the stress tensor,
\begin{equation}
 \frac{\partial\mathcal L_{\mathrm{NM}}}{\partial\lambda}=\frac{(s-1)^2(s+2)}{3\lambda^2},\qquad
  \frac p{\mathcal E}=\frac{1-r}{1+r}=\frac{s^2+s+1}{2s^2+2s-1}\in\Big(\frac12,1\Big],
\end{equation}
where $\mathcal E=2\tau\dot\ell-\ell$ and $p=\ell$ as in section~\ref{sec:causality}.

\subsection{Singular points and physical interpretation}
\label{sec:NM_sing}

\begin{itemize}
\item $r=0$ (vacuum): $r\simeq(s-1)/2$ is analytic and one-to-one, and $\frac{\partial\mathcal L_{\mathrm{NM}}}{\partial\lambda}\to\tfrac14e^{2\gamma}\tau^2$, which gives the quartic coefficient $\lambda/4>0$ in $\mathcal L=S+\tfrac14\lambda(S^2+P^2)+\dots$ at $\gamma=0$. The weak-field expansion \eqref{eq:FNMexp}
contains odd powers of $\rho\propto|T^\mu{}_\mu|$. These are analytic on the causal domain, where $T^\mu{}_\mu\le0$, so the cusp in $\kappa=r$ is an artefact of the coordinate, as in section~\ref{sec:q_sing}.
\item $r\to1/3$ ($\rho\to\sqrt{2/5}$): $\mathcal F_{\rm NM}\simeq\frac{3\sqrt2}{160}(1-3r)^{3/2}\simeq\frac{3}{40s^3}$, a square-root-type branch point of the invariants. It corresponds to $s\to\infty$, i.e.\ $\tau\to\infty$, and not to a finite field strength. It only states that the family realises no stress tensor with $p/\rho<\tfrac12$. In this regime $\ell_{\rm NM}\propto\tau^{3/2}$ and $\mathcal L\sim|F|^3$.
\item $1+r^2$ never vanishes for real $r$, so there are no poles.
\end{itemize}
Unlike the members of the family with $q<1$ (section~\ref{sec:q_sing}), there is neither a critical electric field nor a finite-$\tau$ singularity: the only singular behaviour of $\mathcal F_{\rm NM}$ is the asymptotic endpoint $r\to1/3$.


\section{The logarithmic theory}
\label{sec:logarithmic}

\subsection{Exact flow}

Finally, we consider the causal logarithmic theory of
Russo--Townsend\cite{Russo:2024kto}, defined by
\begin{equation}
\ell_{\log}(\tau;\lambda,\gamma)
=
-\frac{1}{\lambda}
\log\!\left(1-e^\gamma\lambda\tau\right).
\end{equation}
With $z=e^\gamma\lambda\tau$ as $z=e^\gamma\lambda\tau$, we have
\begin{equation}
f(z)=-\log(1-z)
\qquad\Longrightarrow\qquad
f'(z)=\frac{1}{1-z}.
\end{equation}
Setting
\begin{equation}
s=1-z,
\end{equation}
the two functions \eqref{ab2} become
\begin{equation}
a(s)=\frac{1-s}{s},
\qquad
b(s)=-\frac{1}{s}+1-\log s.
\end{equation}
The signed invariant variable \eqref{rdef} then reads
\begin{equation}
r(s)
=
1+\frac{s\log s}{1-s}.
\end{equation}
The physical inverse is expressed through the lower real branch $W_{-1}$ of the
Lambert $W$ function~\cite{Corless:1996} (defined by $We^{W}=x$ with $W\le-1$ for $-1/e\le x<0$),
\begin{equation}
s(r)
=
\frac{r-1}
{W_{-1}\!\left((r-1) e^{(r-1)}\right)},
\end{equation}
which is the branch required for the physical interval $0<s<1$. The
principal branch $W_0$ instead gives the trivial root $s=1$, physical
only at the vacuum point $r=0$; at $s=1$ the two branches meet. The
exact irrelevant deformation flow can therefore be written in the
autonomous form
\begin{equation}
\frac{\partial\mathcal L_{\log}}{\partial\lambda}
=
X\,\mathcal F_{\log}(\rho),
\end{equation}
with, in terms of the signed variable $r$,
\begin{equation}
F_{\log}(r)
=
\frac{r\,s(r)}
{4[1-s(r)](1+r^2)}.
\end{equation}
The signed and principal variables are related by
\begin{equation}
r=\frac{\rho}{\sqrt{4-\rho^2}},
\end{equation}
and expanding around the weak-field point $\rho=0$ one obtains
\begin{equation}
\begin{aligned}
\mathcal F_{\log}(\rho)
={}&
\frac{1}{8}
-\frac{\rho}{12}
-\frac{7\rho^{2}}{288}
+\frac{49\rho^{3}}{4320}
+\frac{\rho^{4}}{6480}
\\
&+
\frac{3499\rho^{5}}{4354560}
+\frac{67\rho^{6}}{1360800}
+\mathcal O(\rho^{7}).
\end{aligned}
\end{equation}
This expansion agrees with the perturbative logarithmic flow obtained
previously in \cite{Babaei-Aghbolagh:2025cni}. As for generic $q$, the odd
coefficients do not vanish (it is the $q\to0$ limit of \eqref{eFq} after
multiplication by $q$), so the logarithmic flow contains half-integer
powers of $Y/X$. The physical intervals
are
\begin{equation}
0<s<1,
\qquad
0< z<1,
\qquad
0< r<1,
\qquad
0<\rho<\sqrt{2},
\end{equation}
with the causal branch corresponding to $\lambda>0$ and $\gamma\geq0$.
As in the $q$-deformed and no-$\tau$-maximum cases, the flow is exact,
nonperturbative, and autonomous, with the entire deformation encoded
in the single function $\mathcal F_{\log}(Y/X)$.

\subsection{Compact form and analytic structure}
\label{sec:log_an}

The flow function of the previous subsection can be recast in a form that makes the analytic structure transparent. Define
\begin{equation}
  z(r)=1-\frac{r-1}{W_{-1}\!\left((r-1)\,e^{\,r-1}\right)},
  \qquad
  \frac{\partial\mathcal L_{\log}}{\partial\lambda} = X\,\frac{r-r\,z(r)}{4\left(z(r)+r^{2}z(r)\right)} .
  \label{eq:ZZ1def}
\end{equation}
Set $u\equiv r-1$ and $W\equiv W_{-1}(ue^{u})$. Then $1-z=u/W$ and $z=(W-u)/W$, so
\begin{equation}
  \frac{1-z}{z}=\frac{u}{W-u},
  \qquad
  \frac{\partial\mathcal L_{\log}}{\partial\lambda}=\frac{X}{1+r^{2}}\;\frac{r\,u}{4\,(W-u)} .
\end{equation}
Using $X/(1+r^{2})=4a^{2}$, which follows from \eqref{rdef} and \eqref{X}, we obtain
\begin{equation}
  \boxed{\;
  \frac{\partial\mathcal L_{\log}}{\partial\lambda}=(\tau\dot\ell)^{2}\,\frac{r\,(r-1)}{D(r)},
  \qquad
  D(r)\equiv W_{-1}\!\left((r-1)e^{\,r-1}\right)-(r-1)\; }
  \label{eq:ZZ1res}
\end{equation}
with $r$ given by \eqref{rdef}. On the physical branch $\tau\dot\ell>0$ this reads, explicitly,
\begin{equation}
  \frac{\partial\mathcal L_{\log}}{\partial\lambda}=\frac{(\ell-\tau\dot\ell)(\ell-2\tau\dot\ell)}
  {W_{-1}\!\left(\frac{\ell-2\tau\dot\ell}{\tau\dot\ell}\,
  e^{\frac{\ell-2\tau\dot\ell}{\tau\dot\ell}}\right)
  +\frac{2\tau\dot\ell-\ell}{\tau\dot\ell}}
  \qquad (\tau\dot\ell>0).
\end{equation}

\subsection{Singular points}
\label{sec:log_sing}

All singularities of \eqref{eq:ZZ1res} originate from the Lambert function. The map $u\mapsto ue^{u}$ attains its minimum $-1/e$ at $u=-1$ and tends to $0^{-}$ as $u\to-\infty$. Hence $W_{-1}(ue^{u})$ is real only for $u<0$, i.e.\ $r<1$, and the real $r$-axis divides into three regimes.

\paragraph{(i) $r<0$: the denominator vanishes identically.}
For $u\le-1$ ($r\le0$) the $W_{-1}$ branch returns $W_{-1}(ue^{u})=u$, so $D\equiv0$ and $z\equiv0$. The literal expression \eqref{eq:ZZ1def} is therefore undefined on the entire half-line $r<0$, and the real domain of $\frac{\partial\mathcal L_{\log}}{\partial\lambda}$ is $0<r<1$. In terms of the original variables, $r<0$ means $b>0$, i.e.\ $\ell>\tau\dot\ell$ for $\tau\dot\ell>0$, which is excluded in the causal sector by convexity ($b\le0$, section~\ref{sec:causality}).

\paragraph{(ii) $r=0$: branch point of $W$, removable from the right.}
At $r=0$ the argument of $W$ equals $-1/e$, where the branches $W_{0}$ and $W_{-1}$ meet. Setting $r=\varepsilon>0$ gives
\begin{equation}
  e\,x+1=\frac{\varepsilon^{2}}{2}+\frac{\varepsilon^{3}}{3}+\dots,
\end{equation}
and the standard expansion $W=-1-p+p^{2}/3+\dots$ with $p=\sqrt{2(ex+1)}$ yields
\begin{equation}
  W_{-1}=-1-\varepsilon+O(\varepsilon^{3}),
  \qquad
  D=-2\varepsilon+O(\varepsilon^{3}).
\end{equation}
The apparent $0/0$ cancels:
\begin{equation}
  \frac{\partial\mathcal L_{\log}}{\partial\lambda}=a^{2}\,\frac{\varepsilon(\varepsilon-1)}{-2\varepsilon}+\dots
  \;\xrightarrow[r\to0^{+}]{}\;\frac{(\tau\dot\ell)^{2}}{2}.
  \label{eq:ZZ1r0}
\end{equation}
This coincides with the value of $ \frac{\partial\mathcal L_{GBI}}{\partial\lambda}$ in \eqref{eq:ZZphys} at $\ell=\tau\dot\ell$. Moreover, the solution $w(u)\neq u$ of $we^{w}=ue^{u}$ is analytic through $u=-1$, with $w\simeq-2-u$ locally. Continuing to $r<0$ therefore selects the $W_{0}$ branch, and
\begin{equation}
  \frac{\partial\mathcal L_{\log}}{\partial\lambda}^{\rm cont}=a^{2}\,\frac{r(r-1)}{W_{0}\!\left((r-1)e^{\,r-1}\right)-(r-1)}\qquad(r<0)
\end{equation}
is finite and smooth through $r=0$, in contrast with the literal $W_{-1}$ formula discussed in (i).

\paragraph{(iii) $r\to1^{-}$: logarithmic endpoint.}
As $r\to1^{-}$ the argument $x=ue^{u}\to0^{-}$, and
\begin{equation}
  W_{-1}(x)\simeq\ln(-x)-\ln\!\big(-\ln(-x)\big)\to-\infty,
  \qquad \ln(-x)\simeq\ln(1-r).
\end{equation}
The numerator in \eqref{eq:ZZ1res} vanishes linearly while $D\to-\infty$ logarithmically, so
\begin{equation}
  \frac{\partial\mathcal L_{\log}}{\partial\lambda}\simeq(\tau\dot\ell)^{2}\,\frac{1-r}{\left|\ln(1-r)\right|}\;\longrightarrow\;0^{+}.
  \label{eq:ZZ1r1}
\end{equation}
The function is finite at this point, but $r=1$ is a genuine logarithmic branch point: $\frac{\partial\mathcal L_{\log}}{\partial\lambda}$ is continuous there, its first derivative vanishes, and derivatives of order two and higher diverge. For $r>1$ the argument of $W_{-1}$ is positive and $\frac{\partial\mathcal L_{\log}}{\partial\lambda}$ is not real.

In the interior $0<r<1$ one has $W_{-1}<-1<u$, so $D<0$ strictly and $\frac{\partial\mathcal L_{\log}}{\partial\lambda}$ is smooth and pole-free. The zeros of $1+r^{2}$ at $r=\pm i$ would require $b^{2}=-a^{2}$ and are absent for real $\ell,\tau,\dot\ell$. The point $\tau\dot\ell=0$ corresponds to $r\to\infty$ and lies outside the real domain.

\subsection{Summary in terms of $\ell,\tau,\dot\ell$}

The real, regular domain of $\frac{\partial\mathcal L_{\log}}{\partial\lambda}$ is $\tau\dot\ell>0$ and $0<r<1$, i.e.
\begin{equation}
  0<|\ell-\tau\dot\ell|<|\tau\dot\ell| ,
\end{equation}
which for $\tau\dot\ell>0$ means $0<\ell<2\tau\dot\ell$ with $\ell\neq\tau\dot\ell$. The special points are summarized in Table~\ref{tab:sing}.

\begin{table}[t]
\centering
\begin{tabular}{@{}lccl@{}}
\hline
Point & $\ell$ (for $\tau\dot\ell>0$) & Behaviour of $\frac{\partial\mathcal L_{\log}}{\partial\lambda}$ & Type \\
\hline
$r<0$ & $\ell>\tau\dot\ell$ ($b>0$) & undefined ($D\equiv0$) & identically singular \\
$r=0$ & $\ell=\tau\dot\ell$ & $\to(\tau\dot\ell)^{2}/2$ & removable (from the right) \\
$r=1$ & $\ell=2\tau\dot\ell$ or $\ell=0$ & $\to0^{+}$, as in \eqref{eq:ZZ1r1} & logarithmic branch point \\
$r>1$ & $|\ell-\tau\dot\ell|>|\tau\dot\ell|$ & not real & outside domain \\
\hline
\end{tabular}
\caption{Special points of $\frac{\partial\mathcal L_{\log}}{\partial\lambda}$ in terms of $\ell,\tau,\dot\ell$.}
\label{tab:sing}
\end{table}

\subsection{Logarithmic electrodynamics and the flow operator}
\label{sec:log_phys}

We use the notation of section~\ref{sec:causality}: $a=\tau\dot\ell$, $b=\ell-\tau\dot\ell$, $\mathcal E=2a-\ell$, $p=\ell$ and $\kappa=-b/a\in[0,1)$ in the causal sector. In this section $r=\kappa$, which is the variable of $z(r)$ in section~\ref{sec:log_an}.

Take $q\to0$ in \eqref{eq:ellq} with $\mu=\lambda/q$ fixed and $x\equiv e^\gamma\mu\tau$ (so that $\mu$ and $x$ play the roles of $\lambda$ and $z$ in the previous subsections):
\begin{equation}
  \ell_{\log}(\tau)=-\frac1\mu\ln(1-x),\qquad
  \dot\ell_{\log}=\frac{e^\gamma}{1-x},\qquad
  \ddot\ell_{\log}=\frac{\mu e^{2\gamma}}{(1-x)^2}.
\end{equation}
The causality conditions \eqref{eq:causal} reduce to $\gamma\ge0$ and $\mu>0$, with $0\le x<1$. For $\gamma=0$,
\begin{equation}
  \mathcal L=S+\tfrac12\mu\,(S^2+P^2)+O(F^6).
\end{equation}
In terms of $x$,
\begin{equation}
  \kappa=1-\frac{\ell}{\tau\dot\ell}=1+\frac{(1-x)\ln(1-x)}{x},
  \qquad
  \frac p{\mathcal E}=\frac{1-\kappa}{1+\kappa},
\end{equation}
and $\kappa$ increases monotonically from $0$ at $x=0$ to $1$ at $x=1$. The whole physical window $0\le\kappa<1$ is therefore covered exactly once.

Since $\mathcal L=\ell(\tau)-2U/\dot\ell(\tau)$ and Lemma~1 of section~\ref{sec:exact_flows} holds for $\mu$ as well,
\begin{equation}
  \boxed{\; \frac{\partial\mathcal L_{\log}}{\partial\lambda}=\partial_\mu\mathcal L=\partial_\mu\ell_{\log}\big|_{\tau,\gamma}
  =\frac{\tau\dot\ell-\ell}{\mu}=-\frac{T^\mu{}_\mu}{4\mu}
  =\frac1{\mu^2}\Big[\frac{x}{1-x}+\ln(1-x)\Big]\;}
  \label{eq:ZZ1phys}
\end{equation}
This equals $\lim_{q\to0}q\, \frac{\partial\mathcal L_{q}}{\partial\lambda}$ at $\lambda=q\mu$. It is the dilatation Ward identity for the coupling $\mu$ of logarithmic electrodynamics, and it has the sign of $\mu$.

\paragraph{Origin of the Lambert function.}
Put $y=1-x\equiv e^{-u}$. Then
\begin{equation}
  1-\kappa=\frac{u}{e^u-1}\quad\Longleftrightarrow\quad e^u=1+\frac{u}{1-\kappa},
\end{equation}
which follows from the $O(q)$ terms of \eqref{eq:sqdef} in the limit $q\to0$ (the $O(q^0)$ terms cancel identically). The substitution $v=-(u+1-\kappa)$ turns it into
\begin{equation}
  v\,e^{v}=(r-1)\,e^{\,r-1},\qquad r=\kappa .
\end{equation}
The trivial solution $u=0$ is $v=r-1$. The non-trivial solution is $v=W_{-1}\big((r-1)e^{r-1}\big)$, so
\begin{equation}
  u=-D(r),\qquad D(r)=W_{-1}\big((r-1)e^{r-1}\big)-(r-1),\qquad s_q\to y=e^{D(r)} .
\end{equation}
Using $1/\mu=a(1-\kappa)/u$, equation \eqref{eq:ZZ1phys} becomes
\begin{equation}
   \frac{\partial\mathcal L_{\log}}{\partial\lambda}=(\tau\dot\ell)^2\,\frac{r\,(r-1)}{D(r)},
\end{equation}
which is \eqref{eq:ZZ1res} of section~\ref{sec:log_an}. The function $z(r)$ is $z=1-(r-1)/W_{-1}=(W_{-1}-r+1)/W_{-1}$, so $z=D/(D+r-1)$ is a simple function of the root $y$. Thus $ \frac{\partial\mathcal L_{\log}}{\partial\lambda}$ is the one-loop-like operator whose transcendental structure comes only from solving the constraint $\tau=V+U/\dot\ell^2$ for the logarithmic Lagrangian.

\subsection{Physical meaning of the singular points}

\paragraph{(i) $r=0$: the conformal point, analytic in $x$.}
Here $x\to0$ ($F\to0$) and $\kappa\simeq x/2$. The argument $(r-1)e^{r-1}\to-1/e$ is the branch point of the Lambert function, where the two roots $u=0$ and $u\ne0$ of $e^u=1+u/(1-\kappa)$ collide. This is a property of the coordinate $r$ only: $ \frac{\partial\mathcal L_{\log}}{\partial\lambda}$ as a function of $x$ is analytic at $x=0$, with
\begin{equation}
   \frac{\partial\mathcal L_{\log}}{\partial\lambda}\simeq\tfrac12e^{2\gamma}\tau^2\,(1-r)\;\xrightarrow[r\to0]{}\;\tfrac12(\tau\dot\ell)^2 ,
\end{equation}
which equals the quartic seed coupling $\tfrac12\mu(S^2+P^2)$ differentiated with respect to $\mu$. The removable singularity found in section~\ref{sec:log_sing} is thus a coordinate effect. The continuation of $ \frac{\partial\mathcal L_{\log}}{\partial\lambda}$ to $r<0$ through the $W_0$ branch corresponds to $x<0$, i.e.\ $\mu\tau<0$ or $\mu<0$ at fixed $\tau$. This sign of the quartic coefficient violates \eqref{eq:causal}, so the region $r<0$, where the literal $W_{-1}$ formula is singular ($D\equiv0$), is non-causal and carries no physical information.

\paragraph{(ii) $r\to1^-$: the critical field.}
Here $x\to1$ and $\dot\ell\to\infty$, so the electric field is bounded,
\begin{equation}
  V<V_{\max}=\frac{e^{-\gamma}}{\mu}\qquad\big(E^2<2e^{-\gamma}/\mu\big),
\end{equation}
while magnetic fields are unbounded. At the bound
\begin{equation}
  \mathcal E\simeq\frac{2}{\mu(1-x)}\to\infty,\qquad p=\ell=\frac{1}{\mu}\ln\frac1{1-x}\to\infty,\qquad \frac p{\mathcal E}\to0^+ .
\end{equation}
The pressure diverges only logarithmically, whereas it saturates at a finite value for $q>0$. For the normalised operator, $\varepsilon=1-\kappa$ and $u\simeq\ln(1/\varepsilon)$ give
\begin{equation}
  \frac{ \frac{\partial\mathcal L_{\log}}{\partial\lambda}}{(\tau\dot\ell)^2}=\frac{r(1-r)}{u}\simeq\frac{\varepsilon}{|\ln\varepsilon|}\to0^+ ,
\end{equation}
while the unnormalised $ \frac{\partial\mathcal L_{\log}}{\partial\lambda}\simeq a/\mu\to\infty$. The logarithm is the $q\to0$ limit of the power-law branch point $y^q$ of section~\ref{sec:q_sing}, as the exponent $1/(1-q)$ of $\varepsilon^{1/(1-q)}$ is replaced by $\varepsilon\,|\ln\varepsilon|^{-1}$. Beyond $V_{\max}$ the Lagrangian is complex, and for $r>1$ the argument $(r-1)e^{r-1}$ of $W_{-1}$ is positive, so $ \frac{\partial\mathcal L_{\log}}{\partial\lambda}$ is not real. Equivalently, $r>1$ means $\ell<0$, which is excluded by $\ell(0)=0$ and $\dot\ell\ge1$. The singularity at $r=1$ is thus the critical-field singularity of the logarithmic theory, and it marks the limit of validity of the effective description.

\paragraph{(iii) $r<0$ and $r>1$.}
The region $r<0$ is non-causal (item~(i)), and $r>1$ is outside the domain of the theory (item~(ii)). The point $\tau\dot\ell=0$ would correspond to $r\to\infty$ and is excluded since $\dot\ell\ge1$.

\begin{table}[t]
\centering\small
\begin{tabular}{@{}p{2.1cm}p{3.4cm}p{3.2cm}p{3.0cm}p{3.1cm}@{}}
\hline
Point & Field variables & Stress tensor & Behaviour of $ \frac{\partial\mathcal L_{\log}}{\partial\lambda}$ & Interpretation \\
\hline
$r=0$ & $F\to0$, $x\to0$, $r\simeq x/2$ & $T^\mu{}_\mu\to0$, $p\to\mathcal E$ & $\tfrac12e^{2\gamma}\tau^2$, analytic in $x$ & conformal point, quartic coefficient $\mu/2>0$ \\
$0<r<1$ & $0<x<1$ & $\mathcal E\ge p>0$ & smooth, $D<0$ & causal window of logarithmic electrodynamics \\
$r\to1^-$ & $x\to1$, $V\to V_{\max}=e^{-\gamma}/\mu$ & $\mathcal E\to\infty$, $p\sim\ln\frac1{1-x}$ & $\sim\varepsilon/|\ln\varepsilon|$ (normalised) & critical field, logarithmic branch point \\
$r<0$ & $\mu<0$ continuation ($W_0$) & --- & literal formula singular & non-causal \\
$r>1$ & $\ell<0$ & --- & not real & outside the theory \\
\hline
\end{tabular}
\caption{Singular points of $ \frac{\partial\mathcal L_{\log}}{\partial\lambda}$ in terms of fields, stress tensor and the causal parameters $\gamma\ge0$, $\mu>0$.}
\label{tab:phys-ZZ1}
\end{table}

\section{Commutativity with the root-$T\bar T$ flow}
\label{sec:commutativity}

A one-scale Courant--Hilbert family $\mathcal L(\lambda,\gamma)$ is driven by two stress-tensor flows. The marginal flow \eqref{gamma_flow}, together with \eqref{rinvariant}, is the root-$T\bar T$ flow,
\begin{equation}
  \frac{\partial\mathcal L}{\partial\gamma}=\mathcal R_\gamma\equiv\tau\dot\ell=\tfrac14\sqrt{4X-Y},
  \label{eq:Rgamma}
\end{equation}
since $4X-Y=16a^2$ with $a=\tau\dot\ell>0$. The irrelevant flow \eqref{autonomous_flow} is
\begin{equation}
  \frac{\partial\mathcal L}{\partial\lambda}=\mathcal O_\lambda\equiv X\,\mathcal F\!\left(\frac YX\right).
  \label{eq:Olambda}
\end{equation}
Both operators are autonomous: $\mathcal R_\gamma$ contains no explicit $\lambda$ or $\gamma$, and $\mathcal O_\lambda$ contains neither $\lambda$ nor $\gamma$. They depend on the couplings only through the stress tensor of $\mathcal L(\lambda,\gamma)$. The flows commute in the following sense.

\paragraph{Proposition.}
For every one-scale CH family, with all derivatives taken at fixed $U,V$,
\begin{equation}
  \frac{\partial}{\partial\lambda}\Big(\mathcal R_\gamma[\mathcal L]\Big)
  =
  \frac{\partial}{\partial\gamma}\Big(\mathcal O_\lambda[\mathcal L]\Big)
  =
  \frac{\partial^2\mathcal L}{\partial\lambda\,\partial\gamma}.
  \label{eq:commute}
\end{equation}
Equivalently, the two-parameter family is reached from the Maxwell seed by applying $\mathcal O_\lambda$ and $\mathcal R_\gamma$ in either order, as in figure~\ref{fig:commute}.

\emph{Proof.} Let $\alpha,\beta\in\{\lambda,\gamma\}$ and write $\ell_\alpha=\partial_\alpha\ell|_\tau$, $\dot\ell_\alpha=\partial_\alpha\dot\ell|_\tau$. By Lemma~1, $\partial_\alpha\mathcal L|_{U,V}=\ell_\alpha(\tau)$. Varying the constraint $\tau=V+U/\dot\ell^2$ at fixed $U,V$ gives
\begin{equation}
  \partial_\alpha\tau\big|_{U,V}=-N\,\dot\ell_\alpha,
  \qquad
  N\equiv\frac{2U/\dot\ell^{3}}{1+2U\ddot\ell/\dot\ell^{3}} .
\end{equation}
The mixed derivative is therefore
\begin{equation}
  \partial_\beta\partial_\alpha\mathcal L=\ell_{\alpha\beta}+\dot\ell_\alpha\,\partial_\beta\tau=\ell_{\alpha\beta}-N\,\dot\ell_\alpha\dot\ell_\beta ,
\end{equation}
which is symmetric under $\alpha\leftrightarrow\beta$. Since $\partial_\gamma\mathcal L=\mathcal R_\gamma[\mathcal L]$ and $\partial_\lambda\mathcal L=\mathcal O_\lambda[\mathcal L]$ hold for all $(\lambda,\gamma)$, differentiating the first with respect to $\lambda$ and the second with respect to $\gamma$ gives \eqref{eq:commute}. $\square$

For the one-scale form \eqref{lCH} the ingredients are $\ell_{\lambda\gamma}=z^2f''/\lambda^2$, $\dot\ell_\lambda=e^\gamma zf''/\lambda$, $\dot\ell_\gamma=e^\gamma(f'+zf'')$ and $\ddot\ell=e^{2\gamma}\lambda f''$, so that
\begin{equation}
  \frac{\partial^2\mathcal L}{\partial\lambda\,\partial\gamma}
  =\frac{z^2f''(z)}{\lambda^2}-N\,\frac{e^{2\gamma}z\,f''(z)\,\big(f'(z)+zf''(z)\big)}{\lambda},
  \qquad
  N=\frac{2U/(e^{3\gamma}f'^{\,3})}{1+2U\lambda f''/(e^{\gamma}f'^{\,3})}.
\end{equation}

\begin{figure}[t]
\centering
\begin{tikzcd}[row sep=huge,column sep=huge]
		\mathcal{L}(0,0) \arrow[r, blue, "{\cal O}_{\lambda}" blue] \arrow[d,red,"{\cal R}_{\gamma}" red]
		&|[blue]| \mathcal{L}(\lambda, 0) \arrow[d,red, "{\cal R}_{\gamma}" red] \\
		|[red]|\mathcal{L}(0, \gamma) \arrow[r, blue, "{\cal O}_{\lambda}" blue]
		&|[red!50!blue]|  \mathcal{L}(\lambda, \gamma)
	\end{tikzcd}
\caption{Commutative diagram between the irrelevant operator $\mathcal O_\lambda=X\mathcal F(Y/X)$ and the root operator $\mathcal R_\gamma=\tfrac14\sqrt{4X-Y}$ for a one-scale Courant--Hilbert family, in the form of figure~1 of~\cite{Babaei-Aghbolagh:2024uqp}. For the generalized Born--Infeld family $\mathcal L(\lambda,0)$ is Born--Infeld theory, $\mathcal L_{\rm MM}(\gamma)$ is ModMax and $\mathcal L(\lambda,\gamma)$ is the generalized Born--Infeld theory.}
\label{fig:commute}
\end{figure}

This is the exact, non-perturbative counterpart of the perturbative statement in~\cite{Babaei-Aghbolagh:2024uqp,Babaei-Aghbolagh:2025cni} that the irrelevant and root flows commute. Autonomy is essential: an irrelevant operator with explicit dependence on $\gamma$ would not satisfy \eqref{eq:commute} for the same $\mathcal R_\gamma$.

\section{Conclusion and discussion}
\label{sec:conclusions}

We have studied the irrelevant deformations of self-dual nonlinear
electrodynamics whose Courant--Hilbert function takes the one-scale form
$\ell(\tau;\lambda,\gamma)=\lambda^{-1}f(z)$ with $z=e^\gamma\lambda\tau$.
For any theories the energy--momentum tensor is controlled by the two
combinations $\tau\dot\ell$ and $\ell-\tau\dot\ell$, and the quadratic
invariants $X$ and $Y$ are functions of the single variable $z$.
Eliminating $z$ in favour of $Y/X$ gives the exact flow
\begin{equation}
\frac{\partial\mathcal L}{\partial\lambda}
=
X\,\mathcal F\!\left(\frac{Y}{X}\right),
\end{equation}
which depends neither on $\lambda$, nor on $\gamma$. Together with the universal marginal flow
$\partial_\gamma\mathcal L=\tau\dot\ell=\tfrac14\sqrt{4X-Y}$, it describes
the two-parameter family $\mathcal L(\lambda,\gamma)$ entirely in terms of
the stress tensor. Since both equations hold for the same family, the two
flows commute on one-scale theories. This is shown in
section~\ref{sec:commutativity}: with all derivatives at fixed $U,V$,
$\partial_\lambda(\mathcal R_\gamma[\mathcal L])=\partial_\gamma(\mathcal O_\lambda[\mathcal L])=\partial_\lambda\partial_\gamma\mathcal L$,
which gives the commutative diagram of figure~\ref{fig:commute} exactly,
without a weak-field expansion. The proof relies on the autonomy of both
operators.

The operator content of the flow is encoded in the expansion
$\partial_\lambda\mathcal L=\sum_m C_m Y^{m/2}X^{1-m/2}$: integer powers
of $X$ and $Y$ correspond to even $m$, half-integer powers to odd $m$. The
generalized Born--Infeld theory, which is the member $q=1/2$ of the
$q$-deformed family, is the case in which the expansion truncates, and the
flow is exactly $\tfrac18X-\tfrac1{16}Y$. For the other theories
considered, namely the $q$-deformed family at $q=3/4$, the
no-$\tau$-maximum theory ($q=3/2$ at negative coupling) and the
logarithmic theory ($q\to0$), the odd coefficients are non-zero and
half-integer powers are present. The logarithmic case is the most
transcendental one: the exact flow is expressed through the lower real
branch of the Lambert $W$ function, and its weak-field expansion
reproduces the perturbative flow obtained previously. The agreement of the
weak-field coefficients of all examples with the perturbative results of
the literature provides a check of the exact formulas.

The singular points of the flow functions have a direct physical meaning
in each theory. For the $q$-deformed theories with $q<1$ and for the
logarithmic theory, the endpoint of the range of $Y/X$ is the critical
electric field, where $\dot\ell\to\infty$; the endpoint is a power-law
branch point for the $q$-family and a logarithmic branch point in the
logarithmic theory. For $q>1$, in particular for the no-$\tau$-maximum
theory, there is no critical field and the endpoint of the range
corresponds to $\tau\to\infty$, where the ratio of pressure to energy
density approaches $1/(2q-1)$. The apparent singularity at the
weak-field point (the branch point of the Lambert function for the
logarithmic theory, the cusp in $\kappa$ for odd coefficients) is an
artefact of the variable $\kappa\ge0$ and not a property of the theory. The
sign of the flow is fixed by the sign of $\lambda$ through the dilatation
Ward identity, consistently with the positivity of the quartic coefficient
required by causality.

The construction has several limitations that should be kept in mind.
First, it applies to one-scale CH functions; the flow function
$\mathcal F$ is specific to each family and is therefore not a universal
operator acting on all theories. Second, $\mathcal F$ is defined only on
the range of the map $z\mapsto Y/X$, and it is single-valued only on
branches where this map is invertible. We verified this property
analytically or numerically for the examples treated here, but we did not
establish it for a general $f$. Third, the limit $q\to1$ is degenerate: at
the ModMax point $\ell=e^\gamma\tau$ one has $\ell-\tau\dot\ell\equiv0$, so
$Y/X\equiv0$, the map is not invertible, and the coefficients $C_m(q)$ in
\eqref{eFq} diverge.

Several directions follow naturally. The most immediate one is the inverse
problem: given a flow function $\mathcal F(Y/X)$, can one reconstruct the
CH function $f$, is the solution unique with a Maxwell seed, and which
$\mathcal F$ are compatible with the causality conditions $\dot\ell\ge1$
and $\ddot\ell\ge0$? Answering these questions would turn the present
description into a characterization of the admissible flows. A second
direction is the classification of the one-scale CH functions beyond the
examples studied here, in particular determining which of them give a
globally invertible map $z\mapsto Y/X$ and which require a branchwise
treatment. A third direction is to go beyond one-scale families: the
compatibility of the marginal and irrelevant flows has been established here
(section~\ref{sec:commutativity}) only on theories where both are generated by
the same two-parameter family. Finally, it would be interesting to see whether a similar
description exists for self-dual theories in other dimensions~\cite{Brizio:2026ynw}, where the
stress tensor has more independent invariants, and for supersymmetric
completions~\cite{Kuzenko:2000tg,Ketov:1998ku,Kuzenko:2013gr,Davgadorj:2026nuz}. For the 3-form self-dual field in $D=6$, where only one independent constant is present, our results can be directly generalized using the Ivanov, Nurmagambetov and Zupnik (INZ) formula~\cite{Ivanov:2014nya} discussed in Ref.~\cite{Ferko:2024zth}.

\begin{acknowledgments}
We are grateful to  Song He,  Jorge G. Russo, Dmitri Sorokin, and Roberto Tateo for their interest in this work and the fruitful discussions that followed. We would like to express our special thanks to Dmitri Sorokin for his valuable insights, generous feedback, and encouraging discussions throughout this project. The work of H.B.-A. was conducted as part of the PostDoc Program on {\it Exploring TT-bar Deformations: Quantum Field Theory and Applications,} sponsored by Ningbo University.
\end{acknowledgments}

\bibliographystyle{JHEP}
\bibliography{Ref}
\end{document}